\documentclass[aps,prd,twocolumn,groupedaddress,amssymb,eqsecnum,showpacs,epsfig]{revtex4}
\usepackage{graphicx}
\usepackage{bm}
\usepackage{dcolumn}
\usepackage{amsmath}

\def \lleq {\lower0.9ex\hbox{ $\buildrel < \over \sim$} ~}
\def \ggeq {\lower0.9ex\hbox{ $\buildrel > \over \sim$} ~}

\def \beq  {\begin{equation}}
\def \eeq  {\end{equation}}
\def \ber  {\begin{eqnarray}}
\def \eer  {\end{eqnarray}}

\begin{document}
\newcommand{\newc}{\newcommand}

\newcommand{\ben}{\begin{eqnarray}}
\newcommand{\een}{\end{eqnarray}}
\newc{\be}{\begin{equation}}
\newc{\ee}{\end{equation}}
\newc{\ba}{\begin{eqnarray}}
\newc{\ea}{\end{eqnarray}}
\newc{\bea}{\begin{eqnarray*}}
\newc{\eea}{\end{eqnarray*}}
\newc{\D}{\partial}
\newc{\ie}{{\it i.e.} }
\newc{\eg}{{\it e.g.} }
\newc{\etc}{{\it etc.} }
\newc{\etal}{{\it et al.}}
\newcommand{\nn}{\nonumber}
\newc{\ra}{\rightarrow}
\newc{\lra}{\leftrightarrow}
\newc{\lsim}{\buildrel{<}\over{\sim}}
\newc{\gsim}{\buildrel{>}\over{\sim}}
\title{Matter density perturbations in modified gravity models with arbitrary coupling between matter and geometry.}

\author{Savvas Nesseris}
\email{nesseris@nbi.dk} \affiliation{The Niels Bohr International
Academy, The Niels Bohr Institute, 2100 Copenhagen \O, Denmark}
\date{\today}

\begin{abstract}
We consider theories with an arbitrary coupling between matter and
gravity and obtain the perturbation equation of matter on
subhorizon scales. Also, we derive the effective gravitational
constant $G_{eff}$ and two parameters $\Sigma$ and $\eta$, which
along with the perturbation equation of the matter density are
useful to constrain the theory from growth factor and weak lensing
observations. Finally, we use a completely solvable toy model
which exhibits nontrivial phenomenology to investigate specific
features of the theory. We obtain the analytic solution of the
modified Friedmann equation for the scale factor $a$ in terms of
time $t$ and use the age of the oldest star clusters and the
primordial nucleosynthesis bounds in order to constrain the
parameters of our toy model.
\end{abstract}


\maketitle

\section{Introduction}
There is accumulating observational evidence based mainly on type
Ia supernovae standard candles \cite{SN} that the universe has
entered a phase of accelerating expansion at a recent cosmological
time scale. This expansion implies the existence of a repulsive
factor on cosmological scales which counterbalances the attractive
gravitational properties of matter on these scales. There have
been several theoretical approaches \cite{CST},
\cite{Perivolaropoulos:2006ce} towards the understanding of the
origin of this factor. The simplest such approach assumes the
existence of a positive cosmological constant which is small
enough to have started dominating the universe at recent times.
This model provides an excellent fit to the cosmological
observational data \cite{Komatsu:2008hk} and has the additional
bonus of simplicity and a single free parameter. Despite its
simplicity and good fit to the data, this model fails to explain
why the cosmological constant is so unnaturally small as to come
to dominate the universe at recent cosmological times, a problem
known as the {\it coincidence problem} and there are specific
cosmological observations which differ from its predictions
\cite{Perivolaropoulos:2008ud},\cite{Perivolaropoulos:2008yc}.

In an effort to address this problem two classes of models have
been proposed: The first class assumes that general relativity
(GR) is a valid theory on cosmological scales and attributes the
accelerating expansion to a {\it dark energy} component which has
repulsive gravitational properties due to its negative pressure.
The role of dark energy is usually played by a minimally coupled
to gravity scalar field called {\it quintessence}\cite{quin}.
Alternatively, the role of dark energy can be played by various
perfect fluids (e.g., Chaplygin gas \cite{chapgas}), topological
defects \cite{defectsde}, holographic dark energy
\cite{holographic}, etc. The second class of models attributes the
accelerating expansion to a modification of general relativity on
cosmological scales which converts gravity to a repulsive
interaction at late times and on cosmological scales. Examples of
this class of models include scalar-tensor
theories\cite{BEPS00,stensor},\cite{Capozziello:2007iu}, $f(R)$
modified gravity theories\cite{fRpapers}, braneworld models
\cite{braneworld}, etc.

Of these examples, $f(R)$ theories have received much attention
mainly due to the fact that they can provide a more natural
explanation of the accelerating expansion of the Universe. At the
same time many of these are able to satisfy the solar system
constraints, see, for example, Ref. \cite{Nojiri:2003ft}, and this
is done without the introduction of extra scalar field degrees of
freedom, for a review of viable $f(R)$ gravity models see
\cite{Nojiri:2008nt},\cite{Nojiri:2006ri}, while for the
advantages and the drawbacks of this class of theories see
\cite{Sotiriou:2008ve} and \cite{Sotiriou:2008rp}. Also, the
$f(R)$ theories arise in a wide range of different frameworks: In
quantum field theories in curved spacetime\cite{birrell}, in the
low energy limit of the $D=10$ superstring theory\cite{ON-Mth}, in
the vacuum action for the grand unified theories, etc.

Recently another approach was proposed, first in
\cite{Allemandi:2005qs},\cite{Inagaki:2005qp} and later in
\cite{Bertolami:2007gv}, namely, to add as a nonminimal coupling
to matter a general function of the Ricci scalar. Such nonlinear
couplings of matter were analyzed in the past \cite{Nojiri:2004bi}
in the context of the accelerated expansion of the Universe and in
Refs. \cite{Sotiriou:2008it}, \cite{Sotiriou:2008dh} it was shown
that these theories do not correspond to scalar-tensor gravity and
therefore have a yet unexplored and non-trivial phenomenology.
Furthermore, such theories due to the explicit nonminimal coupling
between the matter content and gravity have a nonvanishing
covariant divergence of the matter energy-momentum tensor
$T_{\mu\nu}$ \cite{Bertolami:2008zh}, \cite{Bertolami:2008ab},
\cite{Harko:2008qz}, which means that there is an energy exchange
between matter and gravity.

In Sec. II we will derive the background equations of motion for a
general Lagrangian density $\frac{1}{2} f_1(R)+G(L_m) f_2(R)$ and
we will discuss some implications about the possible choices for
the matter Lagrangian $L_m$. Having more than one Lagrangians to
describe the same fluid, e.g. $L_m=p$ or $L_m=-\rho$ for
pressureless dust, is not a problem in GR as all choices will give
the same equations of motion, but in theories with a nonminimal
coupling, different choices in general will give different
phenomenology. Nevertheless, we will choose a Lagrangian that
depends only on the density $L_m=L_m(\rho_m)$ in order to keep the
analysis as simple as possible, but quite general at the same
time.

In Sec. III we will derive the matter density perturbation
equations for the case that the matter Lagrangian is a general
function of the density $\rho$. Being able to predict the
evolution of the matter perturbations for such modified gravity
models is necessary in order to compare the theory against the
linear growth data. Such data are in principle able to differentiate between GR
and modified gravity, with the main reasons being the evolution with time
of the gravitational constant \cite{Nesseris:2007pa}, \cite{Nesseris:2006er}.

We will also evaluate two parameters $\eta$ and
$\Sigma=q(1+\eta/2)$ for this particular model, following Refs
\cite{Amendola:2007rr},\cite{Tsujikawa:2007gd} where the first
quantifies the strength of an anisotropic stress while in the
latter, $q$ quantifies the deviation of the gravitational constant
today. These parameters can be used to constrain the theory with
weak lensing observations and if such surveys measure deviations
from GR, for which $(\Sigma,\eta)|_{GR}=(1,0)$ then those two
parameters could be used to differentiate between modified gravity
and dark energy models.

Finally, in Sec. IV we will use a toy model with Lagrangian
density $R/2+(1+\lambda R)G(L_m)$, where $G(L_m)$ is given by
$G(L_m)=A L_m^{n+1}$ and $L_m=-\rho$. This model has the
energy-momentum tensor of pressureless dust, but exhibits
nontrivial phenomenology compared to its counterpart from GR.
Also, we will present an analytical solution to the modified
Friedmann equation and give the scale factor $a$ in terms of the
cosmic time $t$. In order to provide some bounds on the parameters
$\lambda$ and $n$ of our toy model, we will use the following two
observations: the age of the oldest star clusters and the Big Bang
Nucleosynthesis (BBN) bounds on the gravitational constant $G_N$,
which are of the order of $10\%$.

\section{Background equations} The action we will consider is \be
S=\int d^4x\sqrt{-g}\left[\frac{1}{2} f_1(R)+G(L_m) f_2(R)\right]
\label{actionS}\ee where $g$ is the determinant of the metric
$g_{\mu\nu}$ and $f_i=f_i(R)$, $i=1,2$ are arbitrary functions of
the Ricci scalar $R$, while $G(L_m)$ is an arbitrary function of
the matter Lagrangian density $L_m$. We use the metric signature
$(-,+,+,+)$.

Varying the action with respect to the metric $g_{\mu\nu}$ we
obtain the field equations as \ba \left[F_1(R)+2 G(L_m)F_2(R)
\right] R_{\mu\nu}-\frac{1}{2}
f_1(R)g_{\mu\nu}&+&\nn\\+\left[g_{\mu\nu} \Box-\nabla_\mu
\nabla_\nu \right]\left(F_1(R)+2 G(L_m)F_2(R)
\right)&-&\nn\\-g_{\mu\nu}f_2(R) \left[G(L_m)-K(L_m)
L_m\right]&=&\nn\\=f_2(R) K(L_m) T_{\mu\nu} \label{fieldeqs}\ea
where $F_i=\partial f_i /\partial R$, $i=1,2$, $R_{\mu\nu}$ is the
Ricci tensor and $K(L_m)=\partial G(L_m) / \partial L_m$. When
$f_1(R)=f(R)$, $f_2(R)=1$ and $G(L_m)=L_m$ then we recover the
field equations for $f(R)$ gravity.

Finally, we have defined the energy-momentum tensor of the matter
as \be T_{\mu\nu}=-\frac{2}{\sqrt{-g}}\frac{\delta(\sqrt{-g}
L_m)}{\delta g^{\mu\nu}} \label{en-mom-tensor}\ee Assuming that
the matter Lagrangian density $L_m$ depends only on the metric and
not on it's components, we obtain $T_{\mu\nu}=L_m g_{\mu\nu}-2
\partial L_m/\partial g^{\mu\nu}$. In what follows we will assume
that the matter Lagrangian is an arbitrary function of the energy
density $\rho$ only, so that $L_m=L_m(\rho)$. Then the
energy-momentum tensor can be written as \be T^{\mu\nu}=-\rho
\frac{dL_m}{d\rho}u^\mu u^\nu+\left(L_m-\rho\frac{dL_m}{d\rho}
\right)g^{\mu\nu}\label{matter-en-mom}\ee where the four-velocity
$u^\mu=dx^\mu/ds$ satisfies the condition $u^\mu u_\mu=-1$. In
deriving Eq.(\ref{matter-en-mom}) we have used
Eq.(\ref{en-mom-tensor}), the relation $\delta\rho=\frac{1}{2}
\rho(g_{\mu\nu}+u_\mu u_\nu)\delta g^{\mu\nu}$ and the
conservation of the matter current $\nabla_\nu(\rho u^\nu)=0$.
Note that the difference in the sign of the first term between our
Eq.(\ref{matter-en-mom}) and the corresponding one of
Ref.\cite{Harko:2008qz} is due to the difference in the
normalization condition, $u^\mu u_\mu=-1$ for us.

In general the Lagrangian of a perfect fluid (on shell) is equal
to the pressure $L_m=p$, ie for dust it is equal to zero. However,
this is not the only choice as the Lagrangian $L_m=-\rho$ also
reproduces the same energy-momentum tensor $T_{\mu\nu}=\rho u_\mu
u_\nu$. This is not a problem in GR as both choices give the same
equations of motion, but in theories with a nonminimal coupling,
different choices give different phenomenology. This might be a
problem since our choice for the Lagrangian becomes physically
meaningful without having any criteria for choosing it, besides
the observations of course.

This fact has created some debate on the literature lately
\cite{Sotiriou:2008it},\cite{Bertolami:2008zh},\cite{Bertolami:2008ab}
regarding which is is the best choice for the Lagrangian $L_m$ for
the perfect fluid. Nevertheless, we choose a Lagrangian that
depends only on density $L_m=L_m(\rho_m)$ in order to keep the
analysis as simple as possible but quite general at the same time.

In a flat Friedmann-Robertson-Walker metric with a scale factor
$a(t)$, we obtain the background (zero-order) equations: \ba 3
\left[F_1+2 G F_2 \right]H^2&=&f_2 K \rho_m
+\frac{1}{2}\left[\left(F_1+2 G F_2 \right)R-f_1 \right]\nn
\\-f_2(G-K L_m)&-&3H\frac{d}{dt}(F_1+2G F_2)\label{modfried1}\ea \ba &-&2\left[F_1+2 G F_2 \right]
\dot{H}=f_2 K \rho_m \nn +\\&+&\left(\frac{d^2}{dt^2}-H
\frac{d}{dt}\right)\left[F_1+2 G F_2 \right] \label{modfried2}\ea
where $H=\frac{\dot{a}}{a}$, $R=6(2H^2+\dot{H})$ and a dot
represents a derivative with respect to the cosmic time t.

In order to get the evolution equation for the ideal fluid we take
the covariant divergence of Eq. (\ref{fieldeqs}) and by using the
purely geometrical identities \cite{Koivisto:2005yk}: $\nabla^\mu
G_{\mu\nu}=0$ and $(\Box \nabla_\nu-\nabla_\nu
\Box)F=R_{\mu\nu}\nabla^\mu F$ we get\cite{Harko:2008qz} \ba
\nabla^\mu T_{\mu\nu}= (g_{\mu\nu} L_m-T_{\mu\nu})\nabla^\mu
ln(f_2 K)\label{generTmn}\ea

For the case at hand (\ref{generTmn}) gives \ba
\dot{\rho}_m=\frac{-3K[f_2H\rho_m+2F_2(L_m+\rho_m)(4H\dot{H}+\ddot{H})]}{f_2(K+(L_m+\rho_m)K'(L_m)L_m'(\rho_m))}
\label{modenmom} \ea where the prime denotes differentiation with
respect to the particular argument, i.e., $K'(L_m)=\partial
K/\partial L_m$ while  $L_m'(\rho_m)=\partial L_m/\partial
\rho_m$. Obviously, when $G=L_m$ and $f_2(R)=1$ we recover the
usual energy conservation equation of GR and $f(R)$ theories.

\section{Perturbation equations}
We will consider the following perturbed metric with scalar metric
perturbations $\Phi$ and $\Psi$ in a longitudinal gauge: \be
ds^2=-(1+2\Phi)dt^2+a(t)^2(1-2\Psi)\delta_{ij}dx^idx^j
\label{metric} \ee The energy-momentum tensor of the
nonrelativistic matter is decomposed as $T_0^0=-(\rho_m+\delta
\rho_m)$ and $T^0_\alpha=-\rho_m \upsilon_{m,\alpha}$, where
$\upsilon_{m}$ is a velocity potential. The Fourier transformed
perturbation equations are given in the Appendix by Eqs.
(\ref{apper00}), (\ref{apper12}), (\ref{apcont0}) and
(\ref{apcont1}).

Next, we define the gauge invariant matter density perturbation
$\delta_m$ as \be \delta_m\equiv \frac{\delta \rho_m}{\rho_m}+3 H
\upsilon \label{ddm}\ee where \be \upsilon= a \upsilon_m \ee

Following the approach of Ref. \cite{Tsujikawa:2007gd}, we use a
subhorizon approximation under which the leading terms correspond
to those containing $k^2$ and $\delta \rho_m$. Terms that are of
the form $H^2 \Phi$ or $\ddot{\Phi}$ are considered negligible
relative to terms like $(k^2/a^2)\Phi$ for modes well inside the
Hubble radius $(k^2\gg a^2 H^2)$. Under this approximation Eqs.
(\ref{apcont0}),(\ref{apcont1}) and (\ref{ddm}) yield a second
order differential equation for $\delta$, given in the Appendix by
Eq. (\ref{apperturb1}).

Equation (\ref{apperturb1}) compared to the corresponding one of
GR or $f(R)$ theories (see for example \cite{Tsujikawa:2007gd},
\cite{Carloni:2007yv}, \cite{Ananda:2008tx}) has an extra term
which is a function of the matter content of the theory. This
fact, as we will see in what follows, will force Newton's constant
$G_{eff}$ to be dependent on the matter content, something which
is in sharp contrast with $f(R)$ theories, but not scalar-tensor
gravity where $G_{eff}$ may depend on the scalar field $\phi$
through it's nonminimal coupling $F(\phi)$.

Next, we write $\delta F_1$ and $\delta F_2$ as \be \delta F_i=
F_{i,R}~ \delta R \ee where $i=1,2$, the comma denotes derivative
with respect to the Ricci scalar $R$ and $\delta R$, under the
subhorizon approximation, is given by:\be \delta R \simeq
-2\frac{k^2}{a^2}\left(2 \Psi-\Phi\right) \label{dR}\ee

Making these substitutions and using the subhorizon approximation
in Eqs. (\ref{apper00}) and (\ref{apper12}) we get a set of
equations for $\Phi$ and $\Psi$ given by Eqs. (\ref{apeqpp1}) and
(\ref{apeqpp2}). The next step is to express $\Phi$ and $\Psi$ in
terms of $\delta_m$. This can be done by solving the system
(\ref{apeqpp1}) and ({\ref{apeqpp2}) for $\Phi$ and $\Psi$. Doing
so we find $\Phi$ and $\Psi$ under the subhorizon approximation,
given by Eqs. (\ref{appsihor}) and (\ref{apphihor}).

From the equation for $\Phi$ (\ref{apphihor}) we can define a
Poisson equation in the Fourier space and attribute the extra
terms that appear on the right-hand side to an effective
gravitational constant $G_{eff}$. Doing so, we get the
gravitational potential \ba &&\Phi = -4 \pi G_{eff}
\frac{a^2}{k^2} \delta _m \rho _m \label{poisson} \ea where
$G_{eff}$ is given by (\ref{apgeff}).

Hence, the equation of matter perturbations (\ref{apperturb1}) can
be written compactly as \ba && \ddot{\delta}_m+2 H \dot{\delta}_m
-4 \pi (G_{eff}+\beta)\rho_m\delta_m\simeq0 \label{perturb2} \ea
where the parameter $\beta$, after using Eqs. (\ref{appsihor}) and
(\ref{apphihor}), is given by (\ref{apbetafunc}). As it can be
seen from Eq. (\ref{apbetafunc}) in the minimal coupling case
[$f_2(R)=1$ and $G(L_m)=L_m$] we get the expected result
$\beta=0$.

We also define two parameters $\eta$, which characterizes the
strength of the anisotropic stress \be \eta\equiv
\frac{\Phi-\Psi}{\Psi} \label{etadef}\ee and $q$ via
$\frac{k^2}{a^2}\Psi=-4 \pi G_0 q \rho_m\delta_m$, where $G_0$ is
the value of the gravitational constant measured by the solar
system experiments today. Then these two parameters are given by
Eqs (\ref{apeta}) and (\ref{apq}).

In order to confront the modified gravity models with weak lensing
observations it is necessary to calculate the potential
$\Phi_{WL}$ that characterizes the deviation of light rays. Under
the assumptions that photons follow null geodesics and that the
geodesic deviation equation holds, the lensing potential is
$\Phi_{WL}\simeq\Phi+\Psi$; see Ref. \cite{Schimd:2004nq} for a
quick derivation and a discussion. However, in theories with
nonminimal coupling to matter this might not always be the case.
In Ref. \cite{Sotiriou:2008it} it was shown that for theories with
Lagrangian $f_1(R)/2+\left(1+\lambda f_2(R)\right) L_m$,
nonminimal coupling corrections to the Maxwell equations which
would otherwise affect standard optics, can actually affect only
long, compared to the radius of the curvature of spacetime,
wavelengths. When this is not the case, i.e. we consider the
high-frequency limit, photons are transverse and propagate along
null geodesics.

Having this in mind, we define a combination of parameters
$\Sigma=q(1+\eta/2)$, for which \be \Phi_{WL}\simeq\Phi+\Psi=-8
\pi G_0 \frac{a^2}{k^2} \rho_m \delta_m \Sigma \label{lensing}\ee
where the explicit form of $\Sigma$ is given by Eq.
(\ref{apsigma}). Again, in the case of the minimal coupling
[$f_2(R)=1$ and $G(L_m)=L_m$] we get the expected result for
$\Sigma$: \be \Sigma_{f_R}=\frac{1}{8 \pi F G_0} \ee which agrees
with the result of Refs.
\cite{Tsujikawa:2007gd},\cite{Amendola:2007rr} for simple $f(R)$
theories. We should remind the reader here that the formulas
derived for the weak lensing potential have been derived under the
assumption that light travels in null geodesics, which may be the
case only for certain models and in the high-frequency limit.

Therefore, this class of theories may have non-trivial effects on
the weak lensing that could potentially be observable. However,
this would also require one to work out the most general case for
the relation between the lensing potential and the metric
perturbations. Since this is beyond the scope of this analysis, we
have left the most general case for future work.

In order to recover $f(R)$ gravity for all the results we have
mentioned, we only have to set $f_1(R)=f(R)$, $f_2(R)=1$, and
$G(L_m)=L_m$. This means that $F_2(R)=0$, $K(L_m)=1$, and
$K'(L_m)=0$. Then, for example, Eq. (\ref{apgeff}) yields \be
G_{eff}=\frac{1}{8\pi F}
\frac{1+4\frac{k^2}{a^2R}m}{1+3\frac{k^2}{a^2R}m}
\label{GefffR}\ee where \be m\equiv\frac{RF_{,R}}{F}\nn \ee also
the anisotropic parameter $\eta$ defined in (\ref{etadef}) and
given by (\ref{apeta}) becomes \be \eta=\frac{2\frac{k^2}{a^2 R}m}
{1+2\frac{k^2}{a^2R}m} \label{etafR}\ee with both equations,
(\ref{GefffR}) and (\ref{etafR}), being in agreement with the
standard results from $f(R)$ gravity \cite{Tsujikawa:2007gd}.

\section{A toy model}
As an example we will now consider the case where we have the
Lagrangian of GR plus a specific nonminimal coupling of gravity to
pressureless dust. Specifically, for our toy model we will assume
that $f_1(R)=R$, $f_2(R)=1+\lambda R$, $L_m(\rho)=-\rho$ and
$G(L_m)=A L_m^{1+n}$, where $\lambda$, $A$ and $n$ are constants.
From Eq. (\ref{matter-en-mom}) we see that the energy-momentum
tensor then is given by $T_{\mu\nu}=\rho u_\mu u_\nu$ as usual.
The modified Friedmann and continuity equations can easily be
found from Eqs. (\ref{modfried1}) and (\ref{modenmom}). For the
continuity equation we find \ba 3 \Omega (a)+a \Omega '(a)=0
\label{modcontexam1} \ea where the derivatives are written in
terms of the scale factor $a$ and we have defined $\rho(a)\equiv3
H_0^2 \Omega(a)$.

Equation (\ref{modcontexam1}) can as usual be solved for
$\Omega(a)$ and with initial conditions $\Omega(1)=1$ we get \ba
\Omega(a)=\frac{1}{a^3} \label{rho-of-a}\ea

For the modified Friedmann we find, after using (\ref{rho-of-a})
\ba \hspace{-1cm}&& H(a)^2= \frac{(-3)^n H_0^{2 n+2} A}{2 (-1)^n
3^{n+1} A (3 n+2) \lambda H_0^{2 n+2}+a^{3 n+3}}
\label{modfriedexam01}\ea By demanding that $H(1)\equiv H_0$ we
can solve for $A$ and after substituting back to the previous
equation we find \ba \hspace{-1cm}  && H(a)^2= \frac{H_0^2}{a^{3
n+3} \left(1-6 H_0^2 (3 n+2) \lambda \right)+6 H_0^2 (3 n+2)
\lambda } \label{modfriedexam1}\ea Obviously, when $\lambda=0$ and
$n=0$ we get the usual, from GR, Friedmann equation for a matter
dominated universe. If we expand (\ref{modfriedexam1}) for
$\lambda\ll 1$ we get
\ba && H(a)^2\simeq H_0^2 [a^{-3 (n+1)} +\nn \\
&&6 a^{-6 (n+1)} \left(a^{3 n+3}-1\right) H_0^2 (3 n+2) \lambda
+O\left(\lambda ^2\right)] \ea Thus, when $\lambda\ll 1$ such
models can be thought of as a sum of ideal fluids with the first
term being attributed to a matter fluid component with constant
equation of state $w=n$, while the second to a dark energy fluid
with variable equation of state $w=w(z)$.

\begin{figure}[!t]
\centering
\vspace{0cm}\rotatebox{0}{\vspace{0cm}\hspace{0cm}\resizebox{.46\textwidth}{!}{\includegraphics{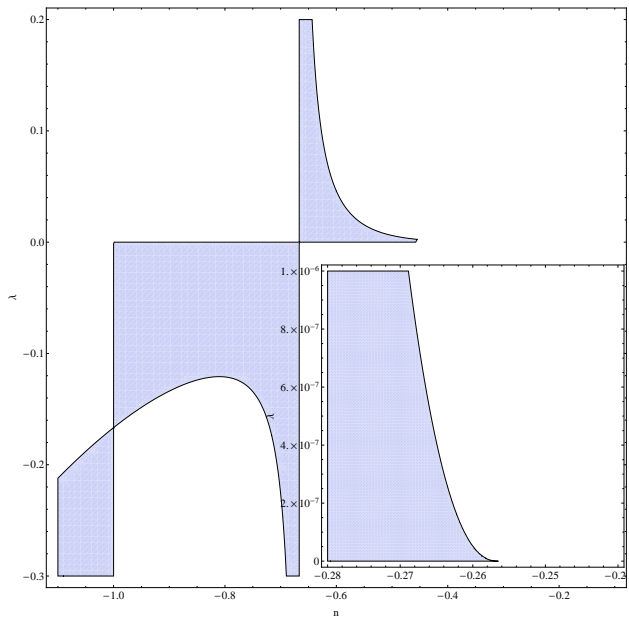}}}
\caption{The allowed parameter space of $(n,\lambda)$ such that
$t_0>12 Gyr$, where $\lambda$ is in units of $1/H_0^2$. The
embedded plot shows the region around $\lambda\ll1$ near the upper
bound for $n$. \label{plot1}}
\end{figure}

We can use Eq.(\ref{modfriedexam1}) to solve for $a(t)$ in terms
of the cosmic time $t$ with the initial condition $a(t_0)=1$,
where $t_0$ is the present time (the age of the universe). Doing
so we find $a(t)$ in implicit form: \be t= \frac{a(t)^{3 n+3} m \,
_3F_2\left(\frac{1}{2},1,1;2,2;-a(t)^{3 n+3} m\right)}{\sqrt{m+1}
(6 n+6)} \ee where $m=\frac{1-6 H_{0}^2 (3 n+2) \lambda }{6
H_{0}^2 (3 n+2) \lambda }$ and $_3F_2$ is a hypergeometric
function. Then the age of the universe is  \be t(a=1)\equiv
t_0=\frac{m \, _3F_2\left(\frac{1}{2},1,1;2,2;-m\right)}{6 H_0
\sqrt{m+1} (n+1)}\label{t0}\ee Demanding that the universe should
be at least old enough to accommodate the oldest star clusters
\cite{Gratton:1997wya}, then it is necessary to have $t_0>12 Gyr$.
Equation (\ref{t0}) also implies that $\lambda$ and $n$ should
satisfy certain conditions in order to have a real value for
$t_0$. Specifically, we find that these conditions are $\lambda>0$
and $n>-\frac{2}{3}$ or $\lambda<0$ and $n<-\frac{2}{3}$. In Fig.
\ref{plot1} we show the allowed parameter space of $(n,\lambda)$
such that $t_0>12 Gyr$. As it can be seen from the embedded plot
there is an upper limit to $n$ that corresponds to $\lambda=0$ and
$t_0=\frac{2}{3 (n+1) H_0}$, with the latter giving $n<-0.256$
after we use the bound $t_0>12Gyr$.

\begin{figure*}[t!]
\centering
\vspace{0cm}\rotatebox{0}{\vspace{0cm}\hspace{0cm}\resizebox{.49\textwidth}{!}{\includegraphics{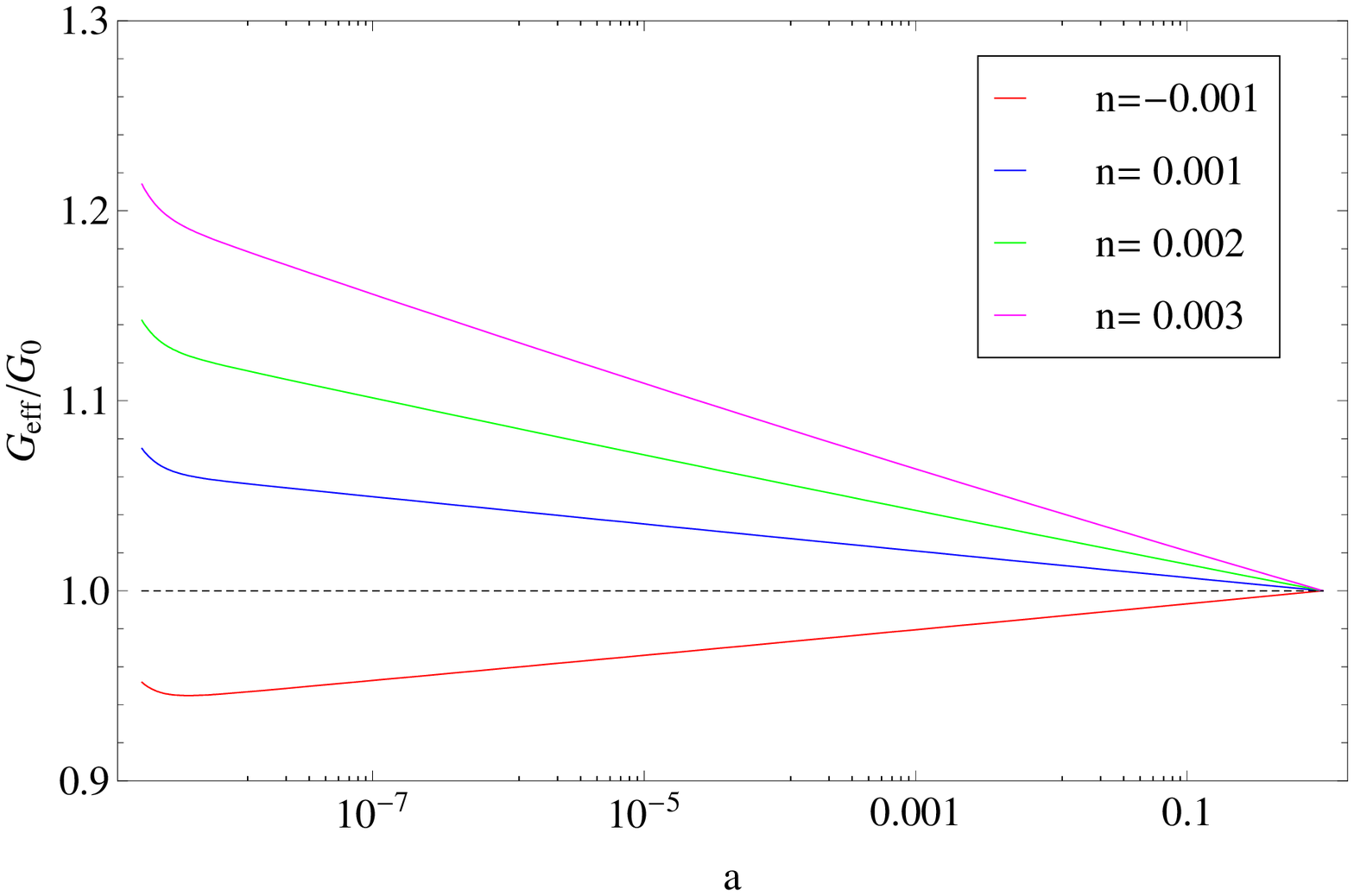}}}
\vspace{0cm}\rotatebox{0}{\vspace{0cm}\hspace{0cm}\resizebox{.49\textwidth}{!}{\includegraphics{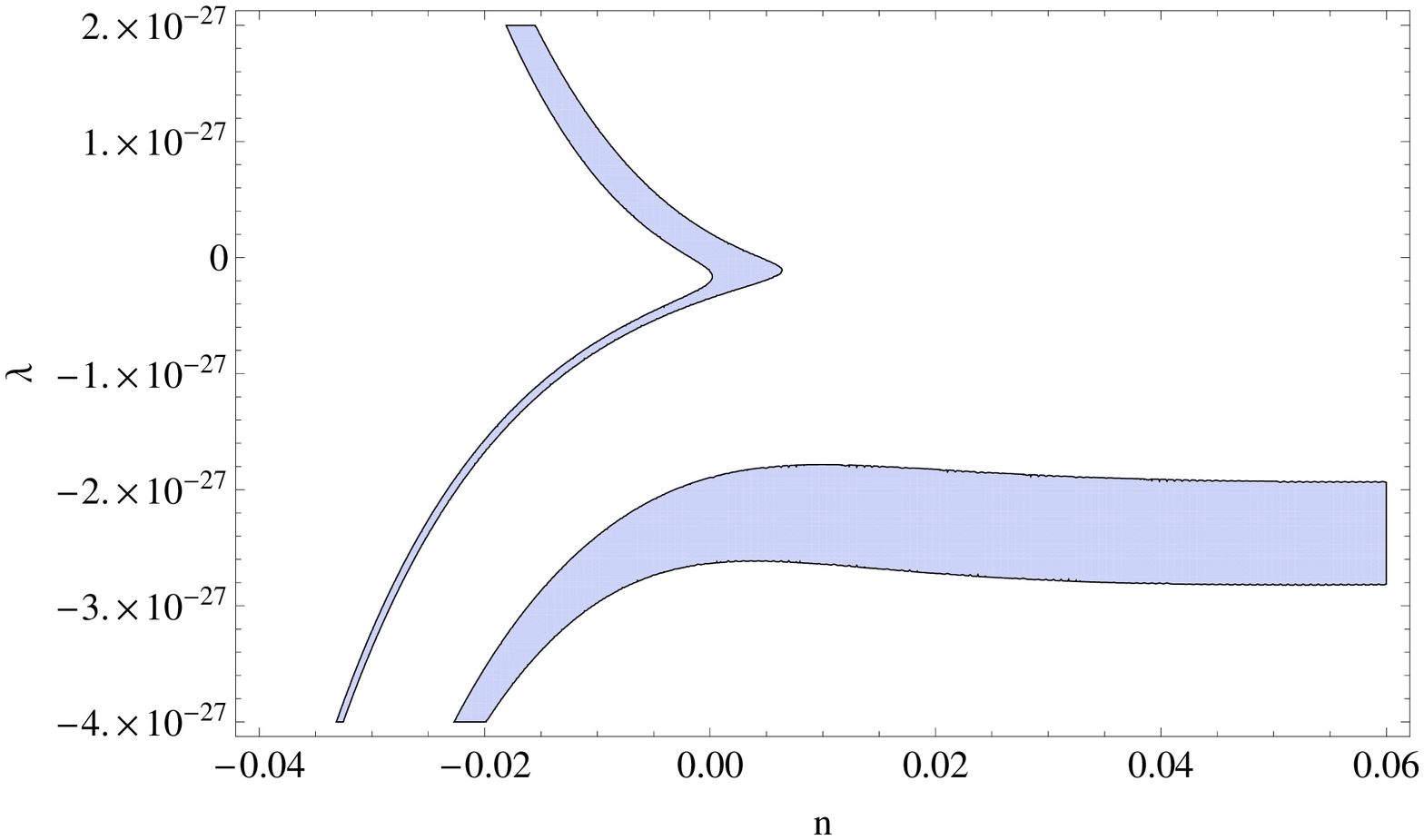}}}
\caption{Left: The evolution of the effective gravitational
constant $G_{eff}/G_0$ for $\lambda=10^{-29}$ and various values
of $n$. Right: The allowed parameter space such that $\frac{\delta
G_{eff}}{G_0}$ is within the $1\sigma$ of the nucleosynthesis
bounds. ($\lambda$ is in units of $1/H_0^2$.) \label{plot2}}
\end{figure*}

Next we will use the primordial nucleosynthesis (BBN) constraints
on the variation of the gravitational constant to further
constrain the parameters $\lambda$ and $n$. The effect of the
variation of $G_{eff}$ can be constrained from BBN to be of the
order of $10\%$, see for example Ref. \cite{Bambi:2005fi} which
gives $\frac{G_{BBN}}{G_0}=1.09\pm^{0.22}_{0.19}$. The effective
gravitational constant $G_{eff}$ can be calculated from
(\ref{apgeff}) by using (\ref{rho-of-a}) and
(\ref{modfriedexam1}). If we set $G_{eff}(a=1)\equiv G_0$ then the
gravitational constant is given by Eq. (\ref{apgefftoy}).

First, we will consider the following two cases, for $\lambda=0$
and $n=0$. In the first case the gravitational constant is simply
given by $G_{eff}=G_0 a^{-3n}$ and by using the BBN constraint of
Ref. \cite{Bambi:2005fi} we find $-0.0018<n<0.0045$ while in the
second case by using a series expansion for small $\lambda$, which
is to be expected from the Solar System tests, we find \be
G_{eff}/G_0=1+\left(2 k^2/H_0^2+\frac{9-2 a
k^2/H_0^2}{a^3}-9\right)H_0^2 \lambda +O\left(\lambda ^2\right)\ee
In this case using the BBN constraint we get $-8.9 \cdot
10^{-29}<\lambda H_0^2<2.8\cdot10^{-28}$. In the general case, the
allowed parameter space, shown in Fig. 2, is rather complicated.
When $n>0$ then $\lambda$ is constrained to be $-2.8 \cdot
10^{-27}<\lambda H_0^2<-1.8 \cdot 10^{-27}$, while when $n<0$ we
cannot give a bound for $\lambda$ in closed form.

It should be noted that while it seems that when we combine the
two constraints there is no allowed parameter space at all for our
model, there are actually values for $(n,\lambda)$ that satisfy
both observational constraints, e.g.
$(n,\lambda)=(-2/3,0.015873)$, but they require extreme
fine-tuning to achieve agreement with the observations and some of
these values exhibit unphysical behavior for $G_{eff}$ at early
times, such as singularities. Therefore we may conclude that our
simple toy model is in some stress with observations as the
allowed parameter space is very small and even then for some of
the allowed parameter values, the model may exhibit unphysical
behavior. This of course does not mean that the whole class of
theories with a nonminimal coupling is problematic, as one may
well choose more suitable forms of the functions $f_1(R)$,
$f_2(R)$ and $G(L_m)$ that satisfy both observational constraints.

\section{Conclusions}
We derived the matter density perturbation equation and the
effective gravitational ``constant" $G_{eff}$ for the action
(\ref{actionS}). Our analysis covers $f(R)$ models with an
arbitrary coupling of the matter content to gravity for any matter
Lagrangian that depends on the fluid density $L_m=L_m(\rho_m)$. In
order to recover the ideal pressureless fluid we chose
$L_m=-\rho$. Having chosen some other dependence for the
Lagrangian our results would be different since the equations of
motion depend explicitly on the choice of the Lagrangian. As we
have mentioned earlier, having more than one Lagrangian to
describe the same fluid, eg $L_m=p$ or $L_m=-\rho$ for
pressureless dust, is not a problem in GR as all choices will give
the same equations of motion, but in theories with a nonminimal
coupling, different choices in general will give different
phenomenology.

We also evaluated two parameters $\eta$, which quantifies the
strength of an anisotropic stress, and $\Sigma=q(1+\eta/2)$, where
$q$ quantifies the deviation of the gravitational constant today.
These parameters can be used to constrain the theory with weak
lensing observations and if such surveys measure deviations from
GR, for which $(\Sigma,\eta)|_{GR}=(1,0)$ then those two
parameters could be used to differentiate between modified gravity
and dark energy models.

Finally, we used a toy model with Lagrangian density
$R/2+(1+\lambda R)G(L_m)$, where $G(L_m)$ is given by $G(L_m)=A
L_m^{n+1}$ and $L_m=-\rho$. This model has the energy-momentum
tensor of pressureless dust, but exhibits nontrivial phenomenology
compared to its counterpart from GR. We presented an analytical
solution to the modified Friedmann equation and gave the scale
factor $a$ in terms of the cosmic time $t$. However, our toy model
is in some stress with observations as the allowed parameter space
is very small. In order to have an old enough universe, so as to
accommodate the oldest observed clusters, the required values for
$\lambda$ are much larger than those allowed by the BBN
constraints (see Figs. 1 and 2). However, this fact should be
interpreted with care since Ref. \cite{Bambi:2005fi} used a simple
modification of the Friedmann equation to derive the BBN
constraints we used. A full analysis of the implications of
modified gravity models with arbitrary coupling between matter and
geometry during nucleosynthesis was beyond the scope of this
analysis and is left for future work.

\section*{Acknowledgements}
The author would like to thank H. Collins, T. Harko, T. Sotiriou
and S. Tsujikawa for useful discussions. This work was supported
by the Niels Bohr International Academy and by the EU FP6 Marie
Curie Research $\&$ Training Network ``UniverseNet"
(MRTN-CT-2006-035863).

\appendix
\section{The perturbation equations} In this Appendix we have
gathered some of the equations that if left in the main text they
would make the analysis difficult to follow.

The Fourier transformed perturbation equations are given by

\begin{widetext}
\ba &&-f_2 K \delta \rho_m - 3 (\delta F_1  +2 G\delta F_2)
\dot{H} +6 F_2 K\dot{H}(2 L_m \Phi+2 \rho_m \Phi-\delta
\rho_m L_m'(\rho_m))\nn\\
&&-f_2 \delta \rho_m (L_m+\rho_m)K'(L_m)L_m'(\rho_m)-3H^2(\delta
F_1+2(F_1 \Phi+G(\delta F_2+2 F_2 \Phi) +F_2
K(-4(L_m+\rho_m)\Phi \nn\\
&& +\delta \rho_m L_m'(\rho_m)))) -3\frac{k^2}{a^2} H^2(\delta F_1
+2(-F_1 \Psi+G(\delta F_2-2 F_2 \Psi)+F_2 K(\rho_m
-L_m(\Phi-2 \Psi)\nn\\
&&+\delta \rho_m L_m'(\rho_m))))  -3 (\dot{F_1}+2 G \dot{F_2}+2
F_2 K L_m'(\rho_m)\dot{\rho_m}) \dot{\Psi}+6 F_2 K
(L_m+\rho_m)\ddot{\Psi} \nn\\
&& + 3 H (-2 \Phi(\dot{F_1}+2 G \dot{F_2}+2 F_2 K L_m'(\rho_m)
\dot{\rho_m})+2F_2 \delta \rho_m L_m'(\rho_m)^2
\dot{\rho_m} K'(L_m)+\dot{\delta F_1}+2 G \dot{\delta F_2}- \nn \\
&& 2 (F_1+2 F_2 G)\dot{\Psi}+ 2 K (\delta F_2 L_m'(\rho_m)+\delta
\rho_m (\dot{F_2} L_m'(\rho_m)+F_2
\dot{\rho_m} L_m''(\rho_m))\nn\\
&& +F_2 (L_m'(\rho_m) \dot{\delta
\rho_m}+(L_m+\rho_m)(\dot{\Phi}+4 \dot{\Psi}))))=0
\label{apper00}\ea

\ba \Phi -\Psi+\frac{{\delta F}_1+2 G\left(L_m\right) \left((\Phi
-\Psi ) F_2+{\delta F}_2\right)+2 K\left(L_m\right) F_2 \delta
\rho _m L_m'\left(\rho _m\right)}{F_1}=0 \label{apper12}\ea

\ba &&-\dot{\delta \rho_m}-3H\delta \rho_m+\rho\left(3 \dot{\Psi}-
\frac{k^2}{a} \upsilon_m\right)= (-6 a^2 (\ddot{H}+4 H \dot{H})
F_2 \delta f_2 (L_m+\rho _m) K^2-f_2 (2 F_2 (12 H^2 \dot{\Phi }
(L_m+\rho_m) a^2\nn \\
&&+3 ((L_m+\rho _m) (2 \Phi \ddot{H}+\dot{H} (3 \dot{\Phi }+4
\dot{\Psi })+\frac{\partial^3\Psi}{\partial t^3})-\ddot{H} \delta
\rho_m (L_m'(\rho_m)+1)) a^2-k^2 (\dot{\Phi }-2 \dot{\Psi
})(L_m+\rho_m)\nn \\
&&+H (-12 \dot{H} \delta \rho _m (L_m'(\rho _m)+1) a^2+(3
(\ddot{\Phi}+4 \ddot{\Psi}) a^2-4k^2 \Psi +2 \Phi (12 \dot{H}
a^2+k^2)) L_m+(3(\ddot{\Phi}+4 \ddot{\Psi}) a^2\nn \\
&&-4 k^2 \Psi +2 \Phi (12 \dot{H}a^2+k^2)) \rho_m))-a^2
\dot{\delta f_2} (L_m+\rho _m)) K^2+a^2 f_2^2 (K (\dot{\delta \rho
_m} (L_m+\rho _m) K'(L_m) L_m'(\rho_m)\nn \\
&&+\dot{\rho_m} \delta \rho _m ((L_m+\rho _m) K''(L_m)L_m'(\rho
_m){}^2+K'(L_m) (L_m'(\rho _m){}^2+L_m'(\rho _m)+(L_m+\rho _m)
L_m''(\rho_m))))\nn \\
&&-\dot{\rho_m} \delta \rho _m (L_m+\rho _m) K'(L_m){}^2 L_m'(\rho
_m){}^2))/\left(a^2 K f_2^2(K+\delta \rho _m K'(L_m) L_m'(\rho
_m))\right) \label{apcont0}\ea

\ba && \Phi \rho _m-a \left(\dot{\rho } \upsilon _m+4 H \rho _m
\upsilon _m+\rho _m \dot{\upsilon }_m\right)=(-12 H^3 K F_2 \rho
_m \upsilon _m a^3+6 H K (H^2+\dot{H}) F_2 \rho _m \upsilon _m
a^3\nn \\
&&+\dot{\rho } f_2 \rho _m \upsilon _m K'(L_m)L_m'(\rho _m) a^3+(K
(L_m \delta f_2+6 a (H^3+3 \dot{H}H+\ddot{H}) F_2
\rho _m \upsilon _m)\nn \\
&&+f_2 L_m\delta \rho _m K'(L_m) L_m'(\rho _m)) a^2)/\left(a^2 f_2
(K+\delta \rho _m K'(L_m) L_m'(\rho _m))\right) \label{apcont1}
\ea

\end{widetext}

Under the subhorizon approximation the differential equation for
the growth of perturbations $\delta$ is \begin{widetext} \ba &&
\ddot{\delta}_m+2 H \dot{\delta}_m
+\frac{k^2\Phi}{a^2}-\frac{k^2}{a^2}((L_m (K-\delta_m \rho _m
K'(L_m) L_m'(\rho _m)) (f_2 \delta_m \rho _m K'(L_m)L_m'(\rho
_m) a^2\nn \\
&&+2 k^2 K (\Phi -2 \Psi ) F_2))/(a^2 K^2 f_2 \rho _m))\simeq0
\label{apperturb1} \ea
\end{widetext}

Using the subhorizon approximation in Eqs. (\ref{apper00}) and
(\ref{apper12}), we get a coupled set of equations for $\Phi$ and
$\Psi$: \ba && \frac{1}{a^4}(-f_2 \delta _m \rho _m (K+(L_m+\rho
_m) K'(L_m) L_m'(\rho _m)) a^4 \nn \\ && -2 k^2 (\Psi F_1+F_2 (2 G
\Psi +K ((\Phi -2 \Psi ) L_m\nn \\ && +\rho _m (\Phi -2 \Psi
-\delta _m L_m'(\rho _m))))) a^2\nn \\ &&+2 (F_{1,R}+2 F_{2,R} G)
k^4 (\Phi -2 \Psi )) =0 \label{apeqpp1}\ea

\ba && (((\Phi -\Psi ) F_1+2 F_2 (G (\Phi -\Psi )+K \delta _m \rho
_m L_m'(\rho _m))) a^2\nn \\ &&+2 (F_{1,R}+2 F_{2,R} G) k^2 (\Phi
-2 \Psi ))/(a^2 F_1) =0 \label{apeqpp2}\ea By solving the system
of equations (\ref{apeqpp1}) and ({\ref{apeqpp2}) for $\Phi$ and
$\Psi$ we find \ba && \Psi= -\frac{a^2}{k^2} \delta _m \rho _m
(f_2 (F_1+2 G F_2) (K+(L_m+\rho _m) \nn \\ && K'(L_m) L_m'(\rho
_m)) a^2+2 k^2 (-K F_2 (F_1+2 F_2 (G\nn \\ &&+K (L_m+\rho _m)))
L_m'(\rho _m)+(F_{1,R}+2 F_{2,R} G)f_2 \nn \\ && (K+(L_m+\rho _m)
K'(L_m)L_m'(\rho_m))))/\nn \\ && (2 (F_1+2 G F_2) ((F_1+F_2 (2 G-K
(L_m+\rho _m))) a^2\nn \\ &&+3 (F_{1,R}+2F_{2,R} G) k^2))
\label{appsihor}\ea and \ba &&\Phi = -\frac{a^2}{k^2}
\delta _m \rho _m (f_2 (F_1+2 G F_2) (K+(L_m+\rho _m)\nn \\
&&K'(L_m) L_m'(\rho _m)) a^2+2 k^2 (K F_2 (F_1+2 F_2 (G-\nn \\ &&2
K
(L_m+\rho _m))) L_m'(\rho _m)+2 F_{1,R} f_2 (K+(L_m+\rho _m)\nn \\
&& K'(L_m) L_m'(\rho _m))+4 F_{2,R} G f_2 (K+(L_m+\rho _m) \nn \\
&& K'(L_m) L_m'(\rho_m))))/(2 (F_1+2 G F_2) ((F_1+F_2 (2 G-\nn \\
&& K (L_m+\rho _m))) a^2+3 (F_{1,R}+2 F_{2,R} G)
k^2))\label{apphihor} \ea

By using (\ref{apphihor}) to define a Poisson equation in the
Fourier space we get $G_{eff}$

\ba &&G_{eff} = \frac{1}{8 \pi} (f_2 (F_1+2 G F_2) (K+(L_m+\rho
_m)\nn \\ &&K'(L_m) L_m'(\rho _m)) a^2+2 k^2 (K F_2 (F_1+2 F_2
(G-\nn \\ &&2 K (L_m+\rho _m))) L_m'(\rho _m)+2 F_{1,R} f_2
(K+(L_m+\rho _m)\nn \\ && K'(L_m) L_m'(\rho _m))+4 F_{2,R} G f_2
(K+(L_m+\rho _m) \nn \\ && K'(L_m) L_m'(\rho_m))))/((F_1+2 G F_2)
((F_1+F_2 (2 G-\nn \\ && K (L_m+\rho _m))) a^2+3 (F_{1,R}+2
F_{2,R} G) k^2))\label{apgeff} \ea

The parameter $\beta$, which corresponds to the last term of Eq.
(\ref{apperturb1}), after using Eqs. (\ref{appsihor}) and
(\ref{apphihor}), is given by \ba && \beta=\frac{1}{4
\pi}\frac{k^2}{a^2} L_m (f_2 (F_2 (K^2+2 G K'(L_m) L_m'(\rho_m))
a^2\nn \\ && +(F_1 a^2+3 (F_{1,R}+2 F_{2,R} G) k^2)K'(L_m)
L_m'(\rho_m))\nn \\ &&
-6 k^2 K^2 F_2^2 L_m'(\rho_m)))/(K f_2 \rho_m (-F_1 a^2\nn \\
&&+F_2 (-2 G+K L_m+K\rho_m) a^2\nn \\ &&-3 (F_{1,R}+2 F_{2,R} G)
k^2)) \label{apbetafunc}\ea

The two parameters $\eta$ and $q$ defined in the main text are
given by \ba && \eta=2 \frac{k^2}{a^2}(2 K
F_2 (F_1+F_2 (2 G-K L_m-K \rho _m)) L_m'(\rho _m)\nn \\
&&+(F_{1,R}+2 F_{2,R} G) f_2 (K+L_m K'(L_m) L_m'(\rho _m)\nn \\
&&+\rho _m K'(L_m)
L_m'(\rho _m)))/(f_2 (F_1+2 G F_2) (K+(L_m+\rho _m) \nn \\
&&K'(L_m) L_m'(\rho_m))+2 k^2/a^2 ((F_{1,R}+2F_{2,R} G) f_2 (K\nn
\\ &&+L_m K'(L_m) L_m'(\rho_m)+\rho _m K'(L_m) L_m'(\rho _m))\nn
\\ &&-K F_2(F_1+2 F_2 (G+K L_m+K \rho_m)) L_m'(\rho_m)))
\label{apeta}\ea and \ba &&
q= \frac{1}{8 \pi G_0}(f_2 (F_1+2 G F_2) (K+(L_m+\rho _m) \nn \\
&& K'(L_m) L_m'(\rho _m)) a^2+2 k^2 (-K F_2 (F_1+2 F_2 (G\nn \\
&&+K (L_m+\rho _m))) L_m'(\rho _m)+F_{1,R} f_2 (K+(L_m+\rho _m)\nn
\\ && K'(L_m) L_m'(\rho _m))+2 F_{2,R} G f_2 (K+(L_m+\rho _m)\nn
\\ && K'(L_m)L_m'(\rho_m))))/((F_1+2 G F_2) ((F_1+F_2 (2 G-\nn
\\ &&K (L_m+\rho _m))) a^2+3 (F_{1,R}+2F_{2,R} G) k^2))
\label{apq}\ea

The explicit form of $\Sigma$, used to confront the modified
gravity models with weak lensing observations, is given by \ba &&
\Sigma=(f_2
(F_1+2 G F_2) (K+(L_m+\rho _m) K'(L_m) L_m'(\rho _m)) a^2\nn \\
&&+3 k^2 ((F_{1,R}+2 F_{2,R} G) K f_2-(L_m+\rho _m)\nn \\
&& (2 K^2 F_2^2-(F_{1,R}+2 F_{2,R} G) f_2 K'(L_m))
L_m'(\rho_m)))/\nn \\
&&(8 G_0 \pi (F_1+2 G F_2) ((F_1+F_2 (2 G-K (L_m+\rho_m)))
a^2\nn \\
&&+3 (F_{1,R}+2 F_{2,R} G) k^2)) \label{apsigma}\ea

In the case of our toy model of Sec. IV, $G_{eff}$ can be
calculated by using Eq. (\ref{apgeff}) and is given by

\begin{widetext}\ba G_{eff}(a)=&&G_0(a (18 (n+1) \lambda H_0^2-1) (-3
\lambda  (9 a^2 (n+1)-8 k^2 (3 n+2) \lambda ) H_0^2 (6 (3 n+2)
\lambda H_0^2-1) a^{3(n+1)}+\nn
\\ &&(a^2-2 k^2 \lambda ) (1-6 (3 n+2) \lambda H_0^2){}^2 a^{6
n+6}+36 (3 n+2) \lambda ^2 (a^2 (3 n+4)-2 k^2 (3 n+2) \lambda )
H_0^4))/\nn
\\ && (((6 (3 n+2) \lambda H_0^2-1) a^{3 n+3}+6 \lambda H_0^2)
(a^{3 n+3} (6 (3 n+2) \lambda H_0^2-1)-6 (3 n+2)\lambda H_0^2){}^2
\nn
\\ &&(1+\lambda (-\frac{2 k^2}{H_0^2}+54 (n+1) (3 n+2) \lambda H_0^2-9
n+3) H_0^2)) \label{apgefftoy}\ea
\end{widetext}

\end{document}